\documentclass[twocolumn,prl,aps,showpacs,superscriptaddress]{revtex4}
\usepackage{xspace,amsmath,amsfonts,amsthm,amssymb,amsbsy,graphicx,color}

\begin{document}

\newtheorem{theo}{Theorem}
\newtheorem{lemma}{Lemma} 

\title{Feedback Control of Non-linear Quantum Systems: a Rule of Thumb}

\author{Kurt Jacobs}

\affiliation{Department of Physics, University of Massachusetts at Boston, 100 Morrissey Blvd, Boston, MA 02125, USA}

\affiliation{Hearne Institute for Theoretical Physics, Department of Physics and Astronomy, Louisiana State University, Baton Rouge, LA 70803, USA}

\author{Austin P. Lund} 

\affiliation{Centre for Quantum Computer Technology, Department of Physics, University of Queensland, St.\ Lucia 4072, Australia.}

\affiliation{Hearne Institute for Theoretical Physics, Department of Physics and Astronomy, Louisiana State University, Baton Rouge, LA 70803, USA}

\begin{abstract}
We show that in the regime in which feedback control is most effective --- when measurements are relatively efficient, and feedback is relatively strong --- then, in the absence of any sharp inhomogeneity in the noise, it is always best to measure in a basis that does not commute with the system density matrix than one that does. That is, it is optimal to make measurements that {\em disturb} the state one is attempting to stabilize. 
\end{abstract}

\maketitle

The manipulation of quantum systems using continuous measurement and feedback control has generated increasing interest in the last few years, due to its potential applications in metrology~\cite{Wiseman95,Berry01}, communication~\cite{Geremia04,Jacobs07} and other quantum technologies~\cite{Ahn02,Hopkins03,Geremia04b,Sarovar04,Steck04}, as well as its theoretical interest, connecting as it does the well-developed field of classical control theory~\cite{Whittle} to fundamental questions regarding the structure of information and disturbance in quantum mechanics~\cite{DJJ,FJ}. 

While the dynamics of closed quantum systems is linear, the introduction of continuous measurement renders the dynanics both non-linear and stochastic~\cite{Habib06}. In certain special cases the resulting evolution can be mapped to a linear classical system driven by Gaussian noise, and as a result classical control theory for linear systems solves the optimal control problem~\cite{BelavkinLQG,DJ,Wiseman05}. However, most quantum systems are not amenable to this technique, and experience from non-linear control theory in classical systems tells us that is unlikely that one can obtain general analytic results for optimal quantum feedback control.  
However, it may be possible to obtain general insights or ``rules of thumb'' that can act as guiding principles in the design of control algorithms. Here we ellucidate one such generaly applicable principle.  

The equation describing the dynamics of a quantum system with arbitrary Hamiltonian $H$ subjected to continuous measurement of an arbitrary observable $X$ is given by the stochastic master equation (SME)~\cite{Brun02,JacobsSteck07} 
\begin{eqnarray}
  d\rho & = &  -(i/\hbar) [H,\rho]dt  - k[X,[X, \rho ]] dt  \nonumber \\ 
           &    &  + 4k[X\rho + \rho X - 2\langle X \rangle \rho] (dr - \langle X \rangle dt)  ,
           \label{eq1}
\end{eqnarray}
where $k$ is the strength of the measurement (loosely the rate at which it extracts information~\cite{DJJ}) and $\rho$ the system density matrix. The observer's measurement record is $r(t)$ where $dr=\langle X \rangle dt + dW/\sqrt{8k}$ and $dW$ is Gaussian white noise satisfying the relation $dW^2=dt$~\cite{WienerIntroPaper}. Feedback control is implemented by continually changing the Hamiltonian in response to the continual stream of measurement results. That is, by making $H(t)$ a function of $r(t')$ for all $t' < t$. We note that the SME is invariant under the transformation $X \rightarrow X+\alpha I$ where $\alpha$ is a real number, so we may always take $X$ to be traceless without loss of generality. In addition, many  natural observables have equi-spaced eigenvalues (e.g. excitation number in a harmonic oscillator). In explicit calculations we will take $X$ to have the eigenspectrum of $J_z$, since this is both traceless and equi-spaced, although we do not expect this choice to have any important effect on the results.  

In feedback control one is usually concerned with stabilizing a quantum system in a given state in the presence of noise, or stabilizing it about a given evolution, and it is this large class of problems that we will consider here. In what follows we will explicitly analyze  the problem of stabilization about a specific target state, although our results will also apply to stabilization about a given evolution, since this can be viewed as the former problem in which the target state changes with time. We will denote the target state by $|0\rangle$, and use as our measure of success the probablity, $P$, that the system will be found in the target state apon making a measurement. The goal of feedback is thus to keep $P$ as close to unity as possible. We will assume minimal constraints on the feedback Hamiltonian and measured observable, since we are interested here in general properties of quantum control, rather than constraints that are applicable to specific systems.  That is, we will assume that the controller has the ability to apply any feedback Hamiltonian $H$ such that $\mbox{Tr}[H^2]/\hbar \leq \mu^2$ for some number $\mu$, and to measure any unitary transformation of $J_z$.  If $|0\rangle$ is an eigenstate of $\rho$, then $P$ is equal to the correponding eigenvalue. 

The first important fact we note is that, given an arbitrary system density matrix $\rho$, with eigenvalues $\{\lambda_i : i=0,\ldots,N\}$ then $P$ is maximized by applying a Hamiltonian to rotate the system so that $|0\rangle$ is the eignevector corresponding to the largest eigenvalue~\cite{DJJ}. Because of the minimal constraints on the  measurement, this means that under the assumption that the noise is homogeneous in the vicinity of the target state, it is always optimal to chose $H(t)$ so that $|0\rangle$ remains as close to an eigenvector of $\rho$ as possible. This is because any unitary transformation of $\rho$ can be compensated for the purposes of feedback by applying the inverse unitary to the measured observable $X$. Thus for homogeneous noise, choosing $H$ to ``eigenvectorize'' the target state has no adverse effects, and thus  optimizes $P$ at all times. 

We will assume now that we are in the regime of good control so that the condition $\Delta \equiv 1 - P \ll 1$ is maintained by the feedback algorithm as the evolution proceeds. This is an important condition because it will allow us to perform an analysis to first order in $\Delta$. We will also assume that the feedback Hamiltonian is sufficiently strong that it is able to keep the target state close to an eigenstate of $\rho$ to good approximation, and that the noise is homogeneous around the target, ensuring that such a procedure is optimal. This will allow us to analyze the performance of the control algorithm purely in terms of the eigenvalues of $\rho$. Generally one would expect to be in the regime of good control whenever $\mu^2 \gg k^2 \gg \beta^2$, where $\beta$ is the average strength of the noise driving the system, and will be defined precisely below. 

We choose $\lambda_0$ to be the largest egenvalue, and from our second assumption $P = \lambda_0$, and $\Delta = \sum_{i=1}^{N-1}\lambda_i$. We also note that the von Neuman entropy $S(\rho)$ and the linear entropy $L(\rho)$ are given by $S(\rho) = L(\rho) = 1 - \mbox{Tr}[\rho^2] = 2\Delta + {\cal O}(\Delta^2)$. We now wish to ask how the basis in which we choose to measure affects our ability to control the system. Once we have chosen the eigen-spectrum of the observable $X$, we are free to choosen any eigenbasis for this observable, all of which are obtained by applying a unitary transformation to $X$ so that the measured observable becomes $\tilde{X} = UXU^\dagger$. We now evaluate the infinitessimal change in the von Neuman entropy due to the measurement, $dS$, for two ``extreme'' choices of the measurement basis. In the first case we chose $X$ so that it commutes with $\rho$, and in the second we choose the observable to be $X_{\mbox{\scriptsize u}} \equiv \tilde{X}$ where $U$ is chosen so that  $X_{\mbox{\scriptsize u}}$ has a basis which is {\em unbiased} with respect to the eigenbasis of $\rho$. This means that every eigenvector of $X_{\mbox{\scriptsize u}}$ has an equal projection of magnitude $1/\sqrt{N}$ onto all the eigenvectors of $\rho$~\cite{Combescure06x}. In this sense the basis of $X_{\mbox{\scriptsize u}}$ is maximally non-commuting with the eigenbasis of $\rho$. We will refer to the measurement of $X$ as a {\em commuting} measurement, and the measurement of $X_{\mbox{\scriptsize u}}$ as an {\em unbiased} measurement. To calculate the infinitesimal change in entropy due to the measurement we use Eq.(\ref{eq1}) with $H=0$ and the fact that $dS = dL = -d(\mbox{Tr}[\rho^2]) = -\mbox{Tr}[2\rho d\rho + (d\rho)^2]$. For the commuting measurement this gives
\begin{eqnarray}
  dS \! & = & \! \sqrt{8 k} S \! \left[ \sum_{j=1}^{N-1} X_j \left( \frac{\lambda_j}{\Delta} \right)  - X_0\right] \! dW + {\cal O}(\Delta^2)
           \label{eq2}
\end{eqnarray}
where the $X_j$ are the eigenvalues of $X$. To first order in $\Delta$ the decrease in entropy caused by the measurement is thus entirely stochastic. In particular, since $\langle dW \rangle = 0$, the average change in the entropy is zero to first order in $\Delta$; the deterministic decrease caused by the measurement is second order in $\Delta$. After each infinitessimal time-step $dt$, we have the oportunity to apply a Hamiltonian so as to transform the system with a unitary $U(dt) = e^{-iHdt/\hbar}$. However, this unitary cannot change $S$, and since $X_{\mbox{\scriptsize c}}$ commutes with $\rho$ the target state remains an eigenstate of $\rho$ after the measurement. As a result Hamiltonian feedback is not able to contribute to the control process, at least in the regime of good control. For feedback to be effective one must wait until the noise has disturbed the system to the extent that $\lambda_0$ is no longer the largest eigenvalue, at which point feedback can be used to swap the eigenvalues and restore this status to $\lambda_0$. 

To calculate the change in entropy resulting from a measurement of $X_{\mbox{\scriptsize u}}$ we proceed as before, but this time note that 1) the diagonal elements of $X_{\mbox{\scriptsize u}} = UXU^\dagger$ are zero due to the fact that $X$ is traceless, and 2) that an unbiased observable $A$ has the property that $\mbox{tr}[A^m \rho^n] = \mbox{tr}[A^m\rho]\mbox{tr}[\rho^n] = \langle A^m \rangle \mbox{tr}[\rho^n]$~\cite{Combes06}. The result is 
\begin{eqnarray}
  dS & = &  - 8 k S  \left[ \sum_{j=1}^{N-1} |X_{\mbox{\scriptsize u}}^{0j}|^2 \left( \frac{\lambda_j}{\Delta} \right) \right] dt + {\cal O}(\Delta^2) .
           \label{eq3}
\end{eqnarray}
In sharp contrast to a measurement that commutes with $\rho$, we see that this time the measurement induces a reduction in the entropy of the system to first order in $\Delta$, and further, that this reduction is purely deterministic. This shows us immediately that unbiased measurements are much more powerful for feedback control than commuting measurements. For measurements that are neither commuting nor unbiased, in general neither the deterministic nor the stochastic terms will vanish. Thus non-commuting measurements on average induce a reduction in entropy to first order in $\Delta$, but the rate of reduction fluctuates randomly. Non-commuting measurements are therefore superior to commuting measurements for feedback control, which is our primary result.  

The above result allows us to derive a particularly simple formula for the performance of a feedback algorithm in the regime of good control employing an unbiased measurement with strong feedback, and in the presence of isotropic noise. To do so we merely need balance the rate of entropy increase due to the noise with the decrease due to the measurement. When the system state is nearly pure, then the rate of entropy production is approximately independent of $\Delta$. For example, for isotropic dephasing noise in a single qubit (whose contribution to the master equation is $d \rho_{\mbox{\scriptsize noise}} = \beta \sum_k[\sigma_k,[\sigma_k,\rho]] dt$ where $k=x,y,z$) the rate of entropy increase is $\dot{S} = 4\beta - {\cal O}(\Delta)$. For an $N$ dimensional system we will correspondingly define the noise strength $\beta$ as $\dot{S}/4$ where $\dot{S}$ is the rate of entropy increase due to the noise. When $k\gg\beta$ the steady state performance of a feedback algorithm using unbiased measurements and strong feedback is thus 
\begin{equation}
  P = 1 - \frac{\beta}{4 k J },  \;\;\;  \mbox{with} \;\;\; J = \sum_{j=1}^{N-1} \frac{ |X_{\mbox{\scriptsize u}}^{0j}|^2 }{(N-1)} .
\end{equation}
Here we have used the fact that on average under isotropic noise all the small $\lambda_i$ are equal to $\Delta/(N-1)$. 

The above analysis immediately raises two questions. Since measurements that do not commute with $\rho$ are more effective at reducing the entropy, one might ask whether it is the unbiased measurements (being the ones that are maximally non-commuting) that provide the optimal entropy reduction. The second question is whether all unbiased measurements are equally effective.  It turns out that while the answer to the first question is yes for qubits (this was shown in~\cite{FJ}), it is not the case for higher dimensional systems. To show this we performed a numerical study of $\langle dS \rangle$ as the measurement basis is transformed from one that commutes with $\rho$ to one that is unbiased. That is, we explored the bases given by the unitary transformations $U(\epsilon) = e^{i\epsilon A}$ for $\epsilon \in [0,1]$ where $U(1)$ corresponds to an unbiased basis. We did this for each of a set of mutually unbiased bases for three and four dimensional systems (specifically we used those given in~\cite{Combescure06x}) and for a random sample of a thousand density matrices $\rho$. We find that in many cases the maximum of $\langle dS(\epsilon) \rangle$ occurs for $\epsilon < 1$. 

To answer the second question we need to examine the matrix elements $X_{\mbox{\scriptsize u}}^{0j}$. If we write the eigenvectors of $X_{\mbox{\scriptsize u}}$ as ${\bf v}_m$, and their respective elements as $v_{mj}$, then the elements of $X_{\mbox{\scriptsize u}}$ are given by $X_{\mbox{\scriptsize u}}^{mj} = \sum_k c_{mj}^k X_k$ where the$X_k$ are the eigenvalues of $X$ and $c_{mj}^k = v_{km}v_{kj}^*$. Since all the elements $v_{mj}$ have the same magnitude, and since the ${\bf v}_m$ are  orthonormal, for every $m\not= j$ the set $\{c_{mj}^k : k = 0,\ldots,N-1\}$ lies on a circle and $\sum_k c_{mj}^k = 0$. Because of this a simple geometrical argument shows that for systems with two or three dimensions, when $X = J_z$ the magnitudes of all the off-diagonal elements of $X_{\mbox{\scriptsize u}}$ are identical and equal to $1/4$ and $1/3$ respectively. Thus for qubits and qutrits all unbiased bases are equally good for feedback control with observables with equispaced eigenvalues. In four dimensions, however, an explicit calculation shows that in general the unbiased bases produce a set of elements $|X_{\mbox{\scriptsize u}}^{0j}|^2$ that are unequal. In this case the effectiveness of the basis can be changed merely by permuting the basis vectors, and thus all unbiased bases are no longer equal for feedback control. At each time-step one would ideally use the basis that maximizes $|dS|$. 

\begin{figure}[t]
\leavevmode\includegraphics[width=1.0\hsize]{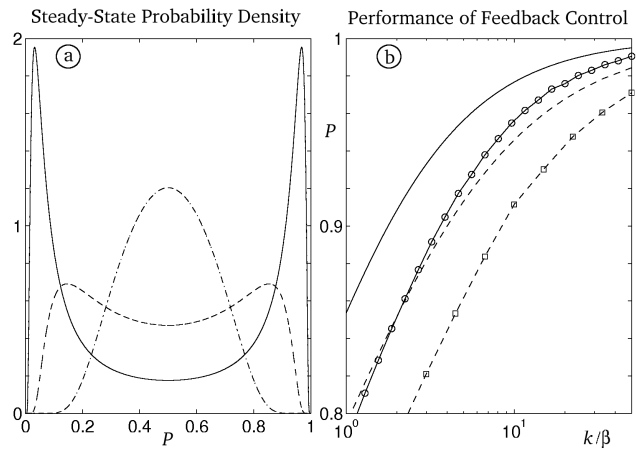}
\caption{In (a) we plot the steady-state probability density for $P$, being the squared overlap with the target state, in the absence of feedback. Solid line: $k/\beta = 2$; Dashed line: $k/\beta = 1/2$; Dash-dot line: $k/\beta = 0.1$. In (b) we plot the average steady-state probability that the system is found in the target state under two feedback algorithms. Solid line: Feedback control using an unbiased measurement with feedback strength $\mu \rightarrow\infty$; Circles:  Same algorithm with $\mu = 100 k$. Dashed: Feedback control using a measurement that commutes with the density matrix for $\mu \rightarrow\infty$. Squares dashed: Same algorithm with $\mu=100k$.} 
\label{fig1}
\end{figure}

We now examine how the above results manifest themselves quantitatively in a concrete application. We consider two feedback control algorithms for a single qubit, the first based on a measurement that commutes with the target state, and the second employing an unbiased measurement. In the first case we are able to obtain a feedback algorithm that is almost certainly optimal for strong feedback, and then compare this with the performance of the second. In both cases we choose the noise to be isotropic so that $d \rho_{\mbox{\scriptsize noise}} = \beta \sum_k[\sigma_k,[\sigma_k,\rho]] dt$.

For the first algorithm we chose to measure the observable $\sigma_z$, the target state to be $|-\rangle$ where $\sigma_z |\pm\rangle = \pm  |\pm\rangle$, and write the density matrix in the eigenbasis of $\sigma_z$. Since both the noise and the measurement of $\sigma_z$ continually destroy the off-diagonal elements of $\rho$, in the absence of feedback we need only consider the diagonal elements of $\rho$, and thus $\rho$ is completely specified by a single dynamical variable, $P$, being the probability that the system is in the target state, or equivalently $z \equiv \langle \sigma_z \rangle = 1 - 2P$. The stochastic master equation then reduces to a stochastic equation for $z$ being $dz = -4 \beta z dt + \sqrt{8k}(1-z^2)dW$.  The Fokker-Planck equation for the probability density of $z$, $p(z,t)$, is $\partial_t p(z,t) = 4\beta \partial_z [ z p(z,t)] + 4 k \partial^2_z [ (1-z^2)^2 p(z,t)]$. We solve this to obtain the steady-state probability density for $z$, which is 
\begin{equation}
  \label{ss-solution}
  p_{\mbox{\scriptsize ss}}(z) = {\cal N}^{-1} \exp \left\{ \beta/[2 k (z^2-1)] \right\}   (z^2-1)^{-2} ,
\end{equation}
where ${\cal N}$ is the normalization. We find that as the measurement strength is increased with respect to the noise power, $p_{\mbox{\scriptsize ss}}(z)$ is increasingly peaked close to the values $z = \pm 1$; as $k$ is increased, the system spends more time in the near-pure states $|\pm\rangle$, and less time in the mixed no-mans-land between the two. This is the regime in which the system exhibits well defined quantum jumps between these states at random intervals. Figure 1(a) shows $p_{\mbox{\scriptsize ss}}(z)$ for three values of $k/\beta$. 

We now wish to add feedback so as to keep the system as close to $|-\rangle$ as possible. Since Hamiltonian evolution can only rotate the state on the Bloch sphere, when $P>0.5$ feedback can only {\em decrease} $P$. We therefore apply a feedback Hamiltonian only when $P<0.5$. In this case a rapid $\pi$ rotation on the Bloch sphere will transform $P\rightarrow P' \approx 1-P$. The key question we must answer is how long we should wait to apply this rotation. If we wait until $P$ is very small, then the rotation will bring the system very close to the target state, but the system will also  spend more time far from the target state. To answer this question we need to solve for the steady-state density in the presence of the feedback algorithm. In implementing the feedback we choose to turn off the measurement when applying the rotation. This both simplifies the analysis and is advisable since the measurement will interfere with the rotation through the quantum Zeno effect. In addition, for strong feedback the rotation takes little time so that removing the measurement has a negligible adverse effect on the entropy.   

In the limit of strong feedback, it turns out that we can include the effect of our feedback algorithm, in which we perform a $\pi$ rotation when $P$ reaches the threshold value $P_{\mbox{\scriptsize T}}$, simply by changing the boundary conditions on the Fokker-Planck equation: with the feedback we now have an absorbing boundary at $z = 1 - 2P_{\mbox{\scriptsize T}} \equiv -\epsilon$, and the point $z = \epsilon$ has an extra probability flux equal to that flowing out the absorbing boundary. The solution for the steady-state density is now 
 \begin{equation}
  \label{ss-solution2}
  p_{\mbox{\scriptsize ss}}^{\mbox{\scriptsize fb}}(z) = \left( A + B \int_0^z \exp\{\beta/[2k(1-x^2)]\}  dx \right) {\cal N}p_{\mbox{\scriptsize ss}}  ,
\end{equation}
where for $z \in [-\epsilon,\epsilon]$ we have $1/A = C$ and $1/B = C \int_{0}^{\epsilon} \exp\{\beta/[2k(1-x^2)]\} dx $, while for $z \in [\epsilon,1]$ we have $1/A = C/2$ and $B = 0$, where $C \equiv \int_{-\epsilon}^{\epsilon} p_{\mbox{\scriptsize ss}}(x) dx +  2\int_{\epsilon}^{1} p_{\mbox{\scriptsize ss}}(x) dx$. Evaluating the integrals numerically we find that the optimal threshold is $\epsilon = 0$, corresponding to $P_{\mbox{\scriptsize T}} = 0.5$. Thus one should apply a $\pi$ rotation to the system as soon as the state crosses the center of the Bloch sphere. We plot the performance of this algorithm, given by the steady-state average success probability, $\langle P \rangle_{\mbox{\scriptsize ss}}$, as a function of $k/\beta$ in Figure 1(b). 

In addition we perform numerical simulations to obtain the performance of this algorithm for a finite feedback Hamiltonian. In this case the threshold is no longer the center of the Bloch sphere: when $|\epsilon|$ is tiny the system will tend to cross the threshold immediately the feedback rotation is complete, invoking a further rotation and effectively freezing the system within the ball $z\in[-\epsilon,\epsilon]$.  We choose a feedback strength of $\mu=100k$, and find numerically the optimal threshold for each value of $k/\beta$. The resulting performance is shown in Figure 1(b). 

We now evaluate the performance of a feedback algorithm that uses an unbiased measurement. In this case at each instant we choose to measure an observable whose basis is unbiased with respect to $\rho$.  In the limit of strong Hamiltonian feedback ($\mu\rightarrow\infty$) we can maintain the direction of the Bloch vector pointing towards the target, so the performance measure $P$ is simply determined by the linear entropy of the state via the relation $L = 2P(1-P)$. In this case we can obtain a simple analytic expression for the performance. The rate of increase of $L$ due to the noise is $\dot{L} = 4\beta (1 - 2L)$ and the rate of decrease due to the measurement is $\dot{L} = -8k L$. The steady-state value of $L$ is the point at which these rates cancel, and is thus $L_{\mbox{\scriptsize ss}} = (1/2)(\beta/(k + \beta))$. This gives $P_{\mbox{\scriptsize ss}} = 0.5 [1 + \sqrt{k/(k+\beta)}]$. To give an example of the performance with a specific finite feedback strength we also perform a numerical simulation with $\mu = 100k$. We plot $P_{\mbox{\scriptsize ss}}$ as a function of $k/\beta$ in Figure 1(b) and compare it to that achieved with the commuting feedback algorithm above. As expected the algorithm employing the unbiased measurement significantly outperforms the algorithm that uses the commuting measurement. 

{\em Acknowledgements:} This work was supported by The Hearne Institute
for Theoretical Physics, The National Security Agency, The Army
Research Office, The Disruptive Technologies Office and the Australian Research 
Council.  


\begin{thebibliography}{21}
\expandafter\ifx\csname natexlab\endcsname\relax\def\natexlab#1{#1}\fi
\expandafter\ifx\csname bibnamefont\endcsname\relax
  \def\bibnamefont#1{#1}\fi
\expandafter\ifx\csname bibfnamefont\endcsname\relax
  \def\bibfnamefont#1{#1}\fi
\expandafter\ifx\csname citenamefont\endcsname\relax
  \def\citenamefont#1{#1}\fi
\expandafter\ifx\csname url\endcsname\relax
  \def\url#1{\texttt{#1}}\fi
\expandafter\ifx\csname urlprefix\endcsname\relax\def\urlprefix{URL }\fi
\providecommand{\bibinfo}[2]{#2}
\providecommand{\eprint}[2][]{\url{#2}}

\bibitem[{\citenamefont{Wiseman}(1995)}]{Wiseman95}
\bibinfo{author}{\bibfnamefont{H.~M.} \bibnamefont{Wiseman}},
  \bibinfo{journal}{{Phys.\ Rev.\ Lett.}} \textbf{\bibinfo{volume}{75}},
  \bibinfo{pages}{4587} (\bibinfo{year}{1995}).

\bibitem[{\citenamefont{Berry et~al.}(2001)\citenamefont{Berry, Wiseman, and
  Breslin}}]{Berry01}
\bibinfo{author}{\bibfnamefont{D.~W.} \bibnamefont{Berry}},
  \bibinfo{author}{\bibfnamefont{H.~M.} \bibnamefont{Wiseman}},
  \bibnamefont{and} \bibinfo{author}{\bibfnamefont{J.~K.}
  \bibnamefont{Breslin}}, \bibinfo{journal}{Phys. Rev. A}
  \textbf{\bibinfo{volume}{63}}, \bibinfo{pages}{053804}
  (\bibinfo{year}{2001}).

\bibitem[{\citenamefont{Jacobs}(2007)}]{Jacobs07}
\bibinfo{author}{\bibfnamefont{K.}~\bibnamefont{Jacobs}},
  \bibinfo{journal}{Quant. Information Comp.} \textbf{\bibinfo{volume}{7}},
  \bibinfo{pages}{127} (\bibinfo{year}{2007}).

\bibitem[{\citenamefont{Geremia}(2004)}]{Geremia04}
\bibinfo{author}{\bibfnamefont{J.~M.}~\bibnamefont{Geremia}},
  \bibinfo{journal}{Phys. Rev. A} \textbf{\bibinfo{volume}{70}},
  \bibinfo{pages}{062303} (\bibinfo{year}{2004}).

\bibitem[{\citenamefont{Ahn et~al.}(2002)\citenamefont{Ahn, Doherty, and
  Landahl}}]{Ahn02}
\bibinfo{author}{\bibfnamefont{C.}~\bibnamefont{Ahn}},
  \bibinfo{author}{\bibfnamefont{A.~C.} \bibnamefont{Doherty}},
  \bibnamefont{and} \bibinfo{author}{\bibfnamefont{A.~J.}
  \bibnamefont{Landahl}}, \bibinfo{journal}{Phys. Rev. A}
  \textbf{\bibinfo{volume}{65}}, \bibinfo{pages}{042301}
  (\bibinfo{year}{2002}).

\bibitem[{\citenamefont{Hopkins et~al.}(2003)\citenamefont{Hopkins, Jacobs,
  Habib, and Schwab}}]{Hopkins03}
\bibinfo{author}{\bibfnamefont{A.}~\bibnamefont{Hopkins}},
  \bibinfo{author}{\bibfnamefont{K.}~\bibnamefont{Jacobs}},
  \bibinfo{author}{\bibfnamefont{S.}~\bibnamefont{Habib}}, \bibnamefont{and}
  \bibinfo{author}{\bibfnamefont{K.}~\bibnamefont{Schwab}},
  \bibinfo{journal}{Phys. Rev. B} \textbf{\bibinfo{volume}{68}},
  \bibinfo{pages}{235328} (\bibinfo{year}{2003}).

\bibitem[{\citenamefont{Sarovar et~al.}(2004)\citenamefont{Sarovar, Ahn,
  Jacobs, and Milburn}}]{Sarovar04}
\bibinfo{author}{\bibfnamefont{M.}~\bibnamefont{Sarovar}},
  \bibinfo{author}{\bibfnamefont{C.}~\bibnamefont{Ahn}},
  \bibinfo{author}{\bibfnamefont{K.}~\bibnamefont{Jacobs}}, \bibnamefont{and}
  \bibinfo{author}{\bibfnamefont{G.~J.} \bibnamefont{Milburn}},
  \bibinfo{journal}{Phys.\ Rev.\ A} \textbf{\bibinfo{volume}{69}},
  \bibinfo{pages}{052324} (\bibinfo{year}{2004}).

\bibitem[{\citenamefont{Steck et~al.}(2004)\citenamefont{Steck, Jacobs,
  Mabuchi, Bhattacharya, and Habib}}]{Steck04}
\bibinfo{author}{\bibfnamefont{D.~A.}~\bibnamefont{Steck}},
  \bibinfo{author}{\bibfnamefont{K.}~\bibnamefont{Jacobs}},
  \bibinfo{author}{\bibfnamefont{H.}~\bibnamefont{Mabuchi}},
  \bibinfo{author}{\bibfnamefont{T.}~\bibnamefont{Bhattacharya}},
  \bibnamefont{and} \bibinfo{author}{\bibfnamefont{S.}~\bibnamefont{Habib}},
  \bibinfo{journal}{Phys. Rev. Lett.} \textbf{\bibinfo{volume}{92}},
  \bibinfo{pages}{223004} (\bibinfo{year}{2004}).

\bibitem[{\citenamefont{Geremia et~al.}(2004)\citenamefont{Geremia, Stockton,
  and Mabuchi}}]{Geremia04b}
\bibinfo{author}{\bibfnamefont{J.~M.} \bibnamefont{Geremia}},
  \bibinfo{author}{\bibfnamefont{J.~K.} \bibnamefont{Stockton}},
  \bibnamefont{and} \bibinfo{author}{\bibfnamefont{H.}~\bibnamefont{Mabuchi}},
  \bibinfo{journal}{Science} \textbf{\bibinfo{volume}{304}},
  \bibinfo{pages}{270} (\bibinfo{year}{2004}).

\bibitem[{\citenamefont{Whittle}(1996)}]{Whittle}
\bibinfo{author}{\bibfnamefont{P.}~\bibnamefont{Whittle}},
  \emph{\bibinfo{title}{Optimal Control}} (\bibinfo{publisher}{Wiley,
  Chichester}, \bibinfo{year}{1996}).

\bibitem[{\citenamefont{Doherty et~al.}(2001)\citenamefont{Doherty, Jacobs, and
  Jungman}}]{DJJ}
\bibinfo{author}{\bibfnamefont{A.~C.} \bibnamefont{Doherty}},
  \bibinfo{author}{\bibfnamefont{K.}~\bibnamefont{Jacobs}}, \bibnamefont{and}
  \bibinfo{author}{\bibfnamefont{G.}~\bibnamefont{Jungman}},
  \bibinfo{journal}{Phys. Rev. A} \textbf{\bibinfo{volume}{63}},
  \bibinfo{pages}{062306} (\bibinfo{year}{2001}).

\bibitem[{\citenamefont{Fuchs and Jacobs}(2001)}]{FJ}
\bibinfo{author}{\bibfnamefont{C.~A.} \bibnamefont{Fuchs}} \bibnamefont{and}
  \bibinfo{author}{\bibfnamefont{K.}~\bibnamefont{Jacobs}},
  \bibinfo{journal}{Phys. Rev. A} \textbf{\bibinfo{volume}{63}},
  \bibinfo{pages}{062305} (\bibinfo{year}{2001}).

\bibitem[{\citenamefont{Habib et~al.}(2006)\citenamefont{Habib, Jacobs, and
  Shizume}}]{Habib06}
\bibinfo{author}{\bibfnamefont{S.}~\bibnamefont{Habib}},
  \bibinfo{author}{\bibfnamefont{K.}~\bibnamefont{Jacobs}}, \bibnamefont{and}
  \bibinfo{author}{\bibfnamefont{K.}~\bibnamefont{Shizume}},
  \bibinfo{journal}{Phys. Rev. Lett.} \textbf{\bibinfo{volume}{96}},
  \bibinfo{pages}{010403} (\bibinfo{year}{2006}).

\bibitem[{\citenamefont{Belavkin}(1987)}]{BelavkinLQG}
\bibinfo{author}{\bibfnamefont{V.~P.} \bibnamefont{Belavkin}}, in
  \emph{\bibinfo{booktitle}{Information, Complexity and Control in Quantum
  Physics}}, edited by
  \bibinfo{editor}{\bibfnamefont{A.}~\bibnamefont{Blaquiere}},
  \bibinfo{editor}{\bibfnamefont{S.}~\bibnamefont{Diner}}, \bibnamefont{and}
  \bibinfo{editor}{\bibfnamefont{G.}~\bibnamefont{Lochak}}
  (\bibinfo{publisher}{Springer-Verlag, New York}, \bibinfo{year}{1987}).

\bibitem[{\citenamefont{Doherty and Jacobs}(1999)}]{DJ}
\bibinfo{author}{\bibfnamefont{A.~C.} \bibnamefont{Doherty}} \bibnamefont{and}
  \bibinfo{author}{\bibfnamefont{K.}~\bibnamefont{Jacobs}},
  \bibinfo{journal}{Phys.\ Rev.\ A} \textbf{\bibinfo{volume}{60}},
  \bibinfo{pages}{2700} (\bibinfo{year}{1999}).

\bibitem[{\citenamefont{Wiseman and Doherty}(2005)}]{Wiseman05}
\bibinfo{author}{\bibfnamefont{H.~M.} \bibnamefont{Wiseman}} \bibnamefont{and}
  \bibinfo{author}{\bibfnamefont{A.~C.} \bibnamefont{Doherty}},
  \bibinfo{journal}{Phys.\ Rev.\ Lett.} \textbf{\bibinfo{volume}{94}},
  \bibinfo{pages}{070405} (\bibinfo{year}{2005}).

\bibitem[{\citenamefont{Brun}(2002)}]{Brun02}
\bibinfo{author}{\bibfnamefont{T.~A.} \bibnamefont{Brun}},
  \bibinfo{journal}{{Am.\ J.\ Phys.}} \textbf{\bibinfo{volume}{70}},
  \bibinfo{pages}{719} (\bibinfo{year}{2002}).

\bibitem[{\citenamefont{Jacobs and Steck}(2007)}]{JacobsSteck07}
\bibinfo{author}{\bibfnamefont{K.}~\bibnamefont{Jacobs}} \bibnamefont{and}
  \bibinfo{author}{\bibfnamefont{D.}~\bibnamefont{Steck}},
  \bibinfo{howpublished}{Contemporary Physics (in press)}
  (\bibinfo{year}{2007}).

\bibitem[{\citenamefont{Gillespie}(1996)}]{WienerIntroPaper}
\bibinfo{author}{\bibfnamefont{D.~T.} \bibnamefont{Gillespie}},
  \bibinfo{journal}{Am. J. Phys.} \textbf{\bibinfo{volume}{64}},
  \bibinfo{pages}{225} (\bibinfo{year}{1996}).

\bibitem[{\citenamefont{Combescure}(2006)}]{Combescure06x}
\bibinfo{author}{\bibfnamefont{M.}~\bibnamefont{Combescure}},
  \emph{\bibinfo{title}{The mutually unbiased bases revisited}},
  \bibinfo{howpublished}{Eprint: quant-ph/0506090} (\bibinfo{year}{2006}).

\bibitem[{\citenamefont{Combes and Jacobs}(2006)}]{Combes06}
\bibinfo{author}{\bibfnamefont{J.}~\bibnamefont{Combes}} \bibnamefont{and}
  \bibinfo{author}{\bibfnamefont{K.}~\bibnamefont{Jacobs}},
  \bibinfo{journal}{Phys. Rev. Lett.} \textbf{\bibinfo{volume}{96}},
  \bibinfo{pages}{010504} (\bibinfo{year}{2006}).

\end{thebibliography}

\end{document}